\documentclass[preprint,3p]{elsarticle}
\usepackage[utf8]{inputenc}
\usepackage{graphicx}           
\usepackage{amsmath}
\usepackage{amssymb}
\usepackage{xspace}
\usepackage{color}
\usepackage{hyperref}
\newcommand{\nc}{\newcommand}
\nc{\rnc}{\renewcommand}
\nc{\bs}{\boldsymbol}
\rnc{\matrix}[2]{\left[\!\!\begin{array}{#1}
	#2\end{array}\!\!\right]}
\rnc{\vector}[1]{\matrix{c}{#1}}
\nc{\mm}[1]{\boldsymbol{#1}}
\nc{\mms}[1]{\boldsymbol{#1}}
\nc{\real}[1]{\Re\left\{ #1 \right\}}
\nc{\imag}[1]{\Im\left\{ #1 \right\}}
\nc{\dd}{\mathrm{d}}
\nc{\ii}{\mathrm{i}}
\nc{\ee}{\mathrm{e}}
\nc{\inv}{^{-1}} 
\nc{\herm}{^{\mathrm H}}
\nc{\tra}{^{\mathrm T}}
\nc{\pinv}{^{+}}
\nc{\conj}[1]{ \overline{#1} }
\nc{\MM}{\mm M}
\nc{\BB}{\mm B}
\nc{\LL}{\mm L}
\nc{\ZZ}{\mm Z}
\nc{\eye}{\mm I}
\nc{\qq}{\mm q}
\nc{\dq}{\dot{\qq}}
\nc{\ddq}{\ddot{\qq}}
\nc{\fex}{\mm f}
\nc{\fexx}{\fex_{\mathrm{ex}}}
\nc{\fexxh}{\hat{\fex}_{\mathrm{ex}}}
\nc{\fcs}{c}
\nc{\fc}{\mm \fcs}
\nc{\fh}{\hat \fcs}
\nc{\fhnull}{\fh_0}
\nc{\xx}{\mm x}
\nc{\xr}{\tilde{\xx}}
\nc{\vv}{\mm\varphi}
\nc{\eex}{\mm e_{\fcs}}
\nc{\eextra}{\mm e\tra_{\fcs}}
\nc{\zA}{Z_{A}}
\nc{\zAst}{\zA^*}
\nc{\za}{Z_{\mathrm a}}
\nc{\ndof}{N_{\mathrm{dof}}}
\nc{\nsens}{N_{\mathrm{sens}}}
\nc{\nmod}{N_{\mathrm{mode}}}
\nc{\phiex}{\varphi_{\mathrm{ex}}}
\nc{\qtip}{q_{\mathrm{tip}}}
\nc{\qex}{q_{\mathrm{ex}}}
\nc{\qexh}{\hat q_{\mathrm{ex}}}
\nc{\mmod}{m_{\mathrm{mod}}}
\nc{\MOD}[1]{#1}
\nc{\TOMOD}[1]{\textcolor{red}{#1}}
\nc{\COMMENT}[1]{\textcolor{blue}{#1}}
\nc{\ie}{i.\,e.\xspace}
\nc{\eg}{e.\,g.\xspace}
\nc{\etc}{etc.\xspace}
\nc{\cf}{cf.\,}
\nc{\myquote}[1]{`#1'}
\nc{\etal}{et al.\xspace}
\nc{\fabstand}{\,}
\nc{\fp}{\fabstand.}
\nc{\fk}{\fabstand,}
\nc{\tab}[5][tbh]{\begin{table}[#1]\centering\caption{#4\label{tab:#5}}\begin{tabular}{#2}\hline #3 \\ \hline\end{tabular}\end{table}}
\nc{\fig}[4][tbh]{
\begin{figure}[#1]
\centering
\includegraphics[width=#4\textwidth]{#2}
\caption{#3\label{fig:#2}}
\end{figure}}
\nc{\e}[2]{\begin{equation} #1 \label {eq:#2} \end{equation}}
\nc{\est}[1]{\begin{equation*} #1 \end{equation*}}
\nc{\ea}[1]{
\begin{eqnarray}
#1\end{eqnarray}}
\nc{\east}[1]{
\begin{eqnarray*}
#1 \end{eqnarray*}}
\nc{\fref}[1]{{Fig.~\ref{fig:#1}}}
\nc{\frefo}[1]{{\ref{fig:#1}}}
\nc{\frefs}[1]{{Figs.~\ref{fig:#1}}}
\nc{\tref}[1]{{Tab.~\ref{tab:#1}}}
\nc{\trefo}[1]{{\ref{tab:#1}}}
\nc{\trefs}[1]{{Tab.~\ref{tab:#1}}}
\nc{\eref}[1]{{Eq.~(\ref{eq:#1})}}
\nc{\erefo}[1]{(\ref{eq:#1})}
\nc{\erefs}[1]{{Eqs.~(\ref{eq:#1})}}
\nc{\sref}[1]{{Section~\ref{sec:#1}}}
\nc{\srefo}[1]{\ref{sec:#1}}
\nc{\srefs}[1]{{Sections~\ref{sec:#1}}}
\nc{\aref}[1]{{{\ref{asec:#1}}}}
\nc{\arefo}[1]{{\ref{asec:#1}}}
\nc{\arefs}[1]{{{Appendices~\ref{asec:#1}}}}

\makeindex             

\begin{document}

\begin{frontmatter}
\title{
Nonlinear damping quantification from phase-resonant tests under base excitation
}
\author{Florian Müller$^1$}
\author{Lukas Woiwode$^1$}
\author{Johann Gross$^1$}
\author{Maren Scheel$^1$}
\author{Malte Krack$^1$}
\address{$^1$ University of Stuttgart, GERMANY}

\begin{abstract}
The present work addresses the experimental identification of amplitude-dependent modal parameters (modal frequency, damping ratio, Fourier coefficients of periodic modal oscillation).
Phase-resonant testing has emerged as an important method for this task, as it substantially reduces the amount of data required for the identification compared to conventional frequency-response testing at different excitation/response levels.
In the case of shaker-stinger excitation, the applied excitation force is commonly measured in order to quantify the amplitude-dependent modal damping ratio from the phase-resonant test data.
In the case of base excitation, however, the applied excitation force is challenging or impossible to measure.
In this work we develop an original method for damping quantification from phase-resonant tests.
It relies solely on response measurement; it avoids the need to resort to force measurement.
The key idea is to estimate the power provided by the distributed inertia force imposed by the base motion.
We develop both a model-free and a model-based variant of the method.
We validate the developed method first in virtual experiments of a friction-damped and a geometrically nonlinear system, and then in a physical experiment involving a thin beam clamped at both ends via bolted joints.
We conclude that the method is highly robust and provides high accuracy already for a reasonable number of sensors.
\end{abstract}



\end{frontmatter}

\section*{Nomenclature}
\begin{eqnarray*}
t && \text{time}\\
\Omega && \text{fundamental angular oscillation frequency}\\
a && \text{modal amplitude}\\
\theta && \text{modal phase lag}\\
\omega && \text{modal frequency}\\
D && \text{modal damping ratio}\\
\mm b && \text{vector imposing the base motion}\\
\mm e && \text{unit vector}\\
\mm \eta && \text{vector of modal coordinates corresponding to $\mm\Phi_{\mathrm{lin}}$}\\
\vv_h && h\text{-th complex Fourier coefficient of mode shape}\\
\vv=\vv_1 && \text{fundamental complex Fourier coefficient of mode shape}\\
q_{\mathrm b} && \text{base displacement}\\
\qq && \text{vector of generalized coordinates}\\
\mm g && \text{vector of generalized stiffness and damping forces}\\
\mm B_{\mathrm{sens}} && \text{matrix selecting the sensor coordinates}\\
\mm I && \text{identity matrix}\\
\MM && \text{mass matrix}\\
\mm\Phi_{\mathrm{lin}} && \text{matrix containing a set of linear mode shapes as columns}\\
\ndof && \text{number of modeled generalized coordinates}\\
\nsens && \text{number of measured response coordinates}\\
\nmod && \text{number of retained linear modes}\\
H && \text{harmonic truncation order}\\
\dot{\square} && \text{derivative with respect to $t$}\\
\hat{\square} && \text{complex Fourier coefficient}\\
\square^* && \text{complex conjugate}\\
\square\tra && \text{(real) transpose}\\
\square\herm && \text{complex-conjugate (Hermitian) transpose}\\
\square^+ && \text{Moore-Penrose pseudo-inverse} 
\end{eqnarray*}

\section{Introduction}\label{sec:intro}
Normal modes are quintessential in structural dynamics.
The natural frequency (or modal frequency) determines the time scale of the vibration and under what conditions resonance phenomena can be expected.
The sign of the modal damping ratio distinguishes positive damping from self-excitation.
The magnitude of a positive damping ratio tells the engineer how quickly free vibrations decay and how large the vibrations become in the resonant case.
The modal deflection shape represents the spatial distribution of the vibration at a well-separated resonance.
All above statements hold for both the linear and the nonlinear case \cite{kers2009}.
\MOD{
Nonlinearity can have a number causes, such as nonlinear kinematic relations, nonlinear material behavior, nonlinear boundary conditions and nonlinear multi-physical interactions.
The nonlinear character of dissipation is well-supported by theory and experiments on different length scales \cite{Eichler.2011,Zaitsev.2012,Amabili.2017}, and it is in many cases dominated by the frictional interactions in mechanical joints \cite{Brake.2018}.
}
While the modal properties are amplitude-independent in the linear case, they are amplitude-dependent in the nonlinear case.
Hence, one obtains additional characteristic information in the nonlinear case:
\begin{itemize}
\item Is a stiffness nonlinearity present and, if so, is it of hardening/softening type?
\item How do resonance frequencies shift with amplitude?
\item Does the damping in- or decrease or even change its sign with amplitude (indicating potential self-excited limit cycle oscillations)?
\item Do mode localization and/or modal interactions occur?
\end{itemize}
Answers to these questions are of utmost value from an engineering perspective.
\\
Modes do not only characterize the vibration signature in the above described way.
They also simplify the quantitative analysis.
Thanks to the superposition principle and orthogonality properties in the linear case, one can simplify the system to a set of decoupled modal oscillators\footnote{This holds strictly only under additional conditions including modal damping.} (\fref{modalOscillators}a).
Superposition is invalid by definition in the nonlinear case.
But in certain situations of utmost engineering relevance, in particular at resonance, a single Nonlinear Mode dominates the response (\emph{Single-Nonlinear-Mode Theory}) \cite{szem1979,krac2013a,Krack.2021}.
In this case, one can reduce the problem to a single nonlinear modal oscillator (with amplitude-dependent properties; \fref{modalOscillators}b).
\fig[t]{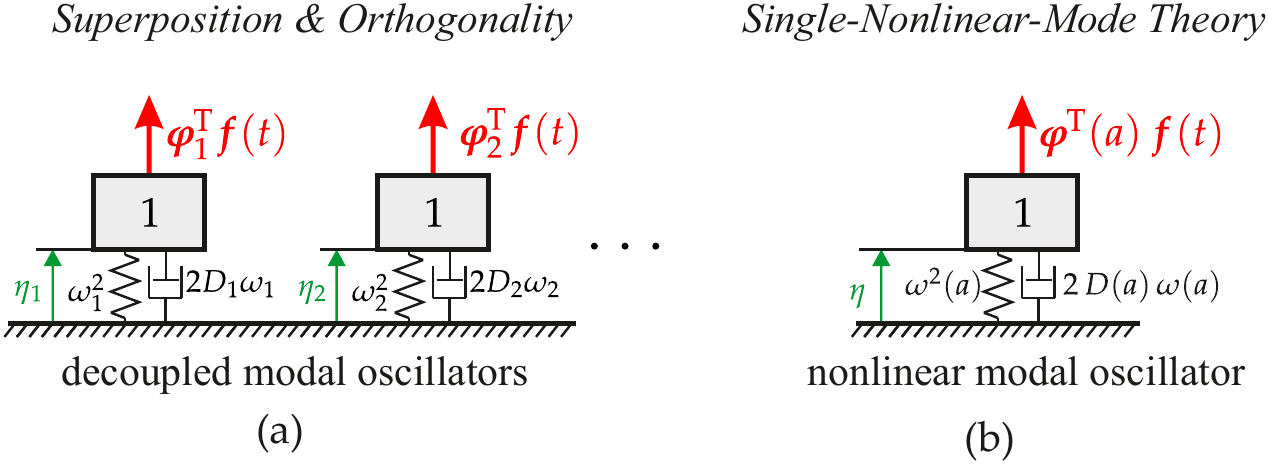}{Modal models: (a) linear case under modal damping, (b) nonlinear case}{0.9}
\\
In some situations a mathematical model of the considered mechanical system is available.
Then a linear modal analysis can be carried out, which is a standard task in every finite element tool.
Relatively well-established computational methods are available for nonlinear modal analysis as well \cite{peet2009,Krack.2015a,Sun.2021,Lacayo.2019b,Balaji.2020,Ponsioen.2018}.
In contrast, methods for experimental nonlinear modal analysis are still under rather active research.
There is some motivation to develop such techniques:
In particular, because the nonlinear modal properties carry characteristic information on the vibration behavior, they are useful quantities for model validation and model updating.
Moreover, one can use the experimentally identified modal properties to feed a nonlinear-modal-oscillator model and make predictions of its response \cite{Peter.2018,Scheel.2018,Karaagacl.2021}.
In this sense, Nonlinear Experimental Modal Analysis is also useful for system identification; and in some cases such a data-driven modeling approach may be preferred over physics-driven ones.
Compared to most alternative methods for nonlinear system identification, an important advantage is that no prior knowledge of the form and location of its local and/or global nonlinearities is needed \cite{Kerschen.2006}.
\\
Several methods have been proposed for Nonlinear Experimental Modal Analysis, see \eg \cite{Peeters.2011,Link.2011,Deaner.2015,Renson.2016,Lacayo.2017,Scheel.2018,Denis.2018,Karaagacl.2021,Kwarta.2022}.
In the present work, we use feedback control to track a sequence of points along the phase resonant backbone curve and identify the modal properties from the periodic steady state.
Using feedback control is more convenient and robust than manual tuning.
Considering the periodic steady state makes the signal processing easy \cite{Dion.2013b,Londono.2015} and a tricky excitation removal is avoided \cite{Mace.2020}.
An alternative to just tracking the backbone curve is to test the frequency response at different response or excitation levels\MOD{, see \eg \cite{Eichler.2011,Zaitsev.2012,Amabili.2017,Karaagacl.2021}}.
Analyzing only the one-dimensional backbone curve instead of a two-dimensional section of the frequency-response surface reduces substantially the amount of required test data.
This is an important point in practice since less data (of essentially the same type) usually means shorter test duration.
As nonlinear vibrations may mean large vibrations and high stresses, longer tests increase the risk of fatigue and significant damage.
Also, long tests may heat up the structure, which, in combination with expansion constraints, can lead to significant changes of the stiffness properties (both linear and nonlinear).

\subsection*{Relevance of base excitation for nonlinear vibration testing}
Besides shaker-stinger excitation, base excitation belongs to the most popular forms of load application for nonlinear vibration testing \cite{Beliveau.1986,ewin1995,Worden.2001}.
Here, the structure is mounted via a (stiff) support frame onto the armature or a slip table of a large shaker (\fref{excitationConfigurations}b-c).
This type of excitation is appropriate if one is interested in the behavior of the structure in the constrained rather than in the totally free configuration.
If the structure under test is actually a component of an assembly, then it is arguably more appropriate to test the structure in a properly constrained configuration.
It should be remarked that an important motivation to test a structure in the free configuration is to avoid uncertainties and nonlinearity associated with constraints at the boundary.
An important motivation for nonlinear vibration testing is to analyze specifically such nonlinear boundary conditions, as introduced in the form of contact interactions in mechanical joints.
This is the first reason why base excitation is (more) relevant for nonlinear vibration testing (than for linear vibration testing).
The second reason is due to the fact that the load is applied in a more distributed way compared to shaker-stinger excitation.
Concentrated loads are associated with stress concentrations near the load application point, which may damage the structure before reaching sufficient vibration level.
In the case of base excitation, the mechanical interface between excitation system (containing exciter and support frame) is larger, as indicated in green color in \fref{excitationConfigurations}, permitting a better stress distribution.
The third and final reason for base excitation is the potential for reducing detrimental exciter-structure interactions.
These interactions may take the form of the resonant force drop \cite{Tomlinson.1979,McConnell.1995}, the occurrence of super- and/or sub-harmonics in the excitation signal, and the divergence of feedback-control loops \cite{Scheel.2018}.
If the structure under test is light compared to the total moving mass (including the shaker armature and the slip table, if applicable), the dynamic forces generated within the structure under test are small compared to the inertia forces of the total moving mass.
Hence, the base motion is not significantly affected by the vibrations of the structure, and it is easier to apply the excitation in the desired form (both in tests without and with feedback control).
Even for relatively small structures, base excitation may be preferred over shaker-stinger excitation, as stingers, load cells and their attachment may significantly deteriorate the dynamic properties \cite{Beliveau.1986}.
\fig[t]{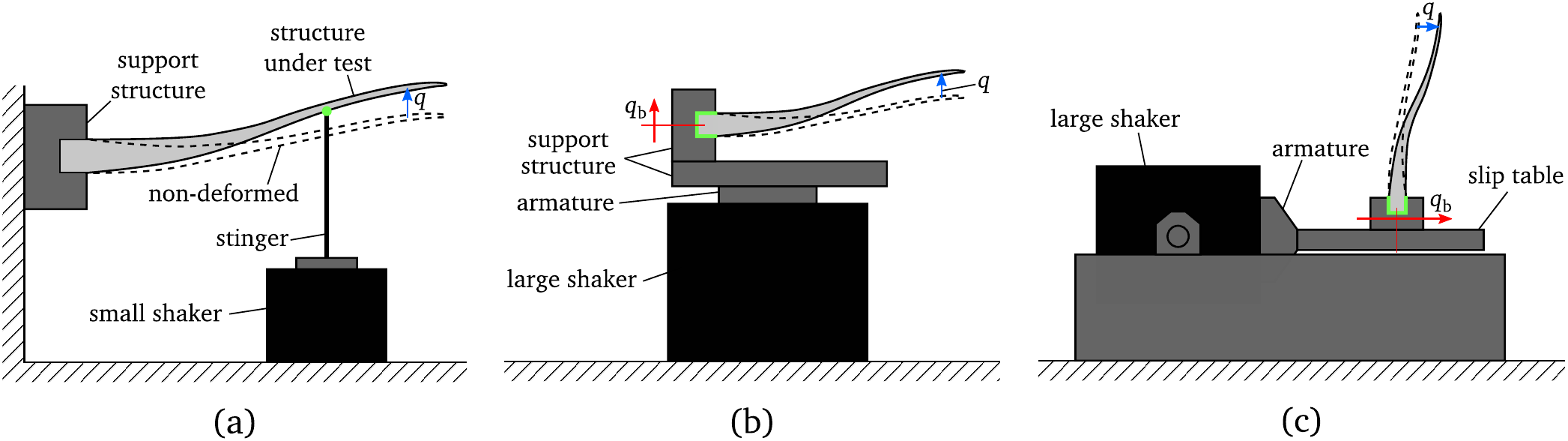}{Typical excitation configurations for vibration tests: (a) application of concentrated force via shaker and stinger, (b) application of base excitation via shaker armature, (c) application of base excitation via slip table}{1.0}

\subsection*{Challenge of damping quantification from phase-resonant tests under base excitation}
For the reasons above, it is desirable to have the capability for Nonlinear Experimental Modal Analysis under base excitation.
However, a challenge lies here in the quantification of the nonlinear damping:
To quantify this from phase-resonant tests, we so far exploited that the power supplied by the excitation force and the dissipated power are equal in average over a vibration period \cite{Scheel.2018}.
Consequently, the applied force must be measured.
An important downside of base excitation is that the applied force cannot be measured directly \cite{ewin1995}.
In \cite{Schwarz.2019}, we built and used a force measurement platform, which was placed between shaker armature and support frame.
We then considered the excitation as external forcing rather than as imposed base motion.
The described work-around leads to additional experimental effort and introduces different sources of uncertainty \cite{Schwarz.2019}.
In particular, the identified damping contains also the dissipation in the support frame, which is not necessarily negligible, as discussed later.
\\
Damping quantification under base excitation is of course also relevant in the linear case.
To the authors' knowledge, there is no method available for modal damping quantification from phase-resonant tests, neither in the nonlinear nor in the linear case.
As in the nonlinear case, it is common to identify modal parameters from frequency response functions (input: base acceleration; output: response relative to base motion) also in the linear case \cite{Beliveau.1986,Vigneron.1987}.

\subsection*{Purpose and outline of the present work}
Motivated by the above described relevance and challenges, the purpose of the present work is to develop a method for nonlinear damping quantification from phase-resonant tests under base excitation.
First, we revisit the method of Nonlinear Experimental Modal Analysis under phase-resonant external forcing (in particular shaker-stinger excitation) in \sref{shakerstinger}.
The theoretical extension to base excitation follows in \sref{extension}, where two variants are developed, a model-free and a model-based one.
The method is applied and assessed using virtual experiments and a physical experiment, the results of which are presented in \sref{resultsVirtual} and \sref{resultsPhysical}, respectively.
Concluding remarks are given in \sref{conclusions}.

\section{Revisiting Nonlinear Experimental Modal Analysis under phase-resonant external forcing\label{sec:shakerstinger}} 
We follow the specific approach developed in \cite{Scheel.2018}.
Although this approach is relatively new, it has already been applied to a number of academic and industrial applications \cite{Peter.2018,Schwarz.2019,Scheel.2020,Scheel.2020b,Abeloos.2022}.
In contrast to the original derivation of the approach in \cite{Scheel.2018}, we reason the approach by Single-Nonlinear-Mode Theory.
More specifically, we seek to solve the \emph{inverse problem}:
Instead of predicting the response for given excitation and modal properties, we require a resonant response and determine the forcing needed to realize this.
It turns out that this strategy permits a quite elegant derivation and is also useful for the extension to base excitation in \sref{extension}.

\subsection{Recap of Single-Nonlinear-Mode Theory\label{sec:NonlinearModeTheory}}
Under steady-state conditions, the governing equation of a single nonlinear modal oscillator reads \cite{szem1979,krac2013a,krac2014a,Krack.2021}
\ea{
\left(~-\Omega^2 + 2D\left(a\right)\omega\left(a\right)\, \ii\Omega + \omega^2\left(a\right)~\right)a = \vv\herm_1\left(a\right)\,\hat{\mm f}_1 \ee^{-\ii\theta}
\fp\label{eq:NMROM}
}
Herein, $a$ and $\theta$ are the real-valued modal amplitude and modal phase, respectively, and $\ii=\sqrt{-1}$ is the imaginary unit.
$\Omega\in\mathbb R_{>0}$ is the fundamental oscillation frequency (here identical to excitation frequency), $\hat{\mm f}_1\in\mathbb C^{\ndof\times 1}$ is the fundamental complex Fourier coefficient of the external forcing.
$\omega(a)$ and $D(a)$ are the amplitude-dependent real-valued modal frequency and damping ratio, respectively, and $\vv_1(a)\in\mathbb C^{\ndof\times 1}$ is the fundamental complex Fourier coefficient of the amplitude-dependent modal deflection shape and $\ndof$ is the number of degrees of freedom (considered finite here).
\eref{NMROM} can be derived by approximating the vibration response in terms of the two-dimensional invariant manifold associated with a given Nonlinear Mode in accordance with the Extended Periodic Motion Concept \cite{Krack.2015a}, and requiring orthogonality of the residual with respect to the fundamental harmonic of the mode, as shown \eg in \cite{krac2014a}.
In \eref{NMROM}, mass normalization is presumed with
\ea{
\vv\herm_1\mm M\vv_1=1\fk \label{eq:massnorm}
}
where $\mm M\in\mathbb R^{\ndof \times \ndof}$ is the symmetric and positive definite mass matrix.
The Fourier coefficients of the associated generalized coordinates and the modal deflection shape are related by:
\ea{
\hat{\mm q}_h = a\ee^{\ii\theta}\vv_h \quad h=0,\ldots,H\fk \label{eq:modeShape}
}
In computational and experimental practice, only a finite harmonic order $H<\infty$ can be considered.

\subsection{Mode isolation using phase-resonant external forcing}
The goal is to achieve and track the resonant response (backbone curve) defined by the resonance condition $\Omega=\omega(a)$.
Substituting these conditions into \eref{NMROM} yields
\ea{
2D\omega^2a\ii = \vv\herm\,\hat{\mm f}_1 \ee^{-\ii\theta} \fp \label{eq:PRTone}
}
Here and in the following, the dependence of $D$, $\omega$ and $\vv$ on $a$ is not explicitly denoted for brevity.
The complex equation \erefo{PRTone} is satisfied if both magnitude and phase of left- and right-hand side agree.
Per definition, $a$ is real and positive.
For positive damping $D>0$, we obtain the condition of phase resonance,
\ea{
\arg\left( \hat{\mm q}\herm_1\,\hat{\mm f}_1 \right) = \frac\pi 2\fk \label{eq:PRTtwo}
}
where we have substituted $\hat{\mm q}_1$ using \eref{modeShape}, which can be directly measured, as opposed to the modal amplitude, phase and mass-normalized deflection shape. 
Suppose that the forcing is applied at a single point via shaker-stinger excitation.
Then we have $\hat{\mm f}_1 = \mm e \hat f$, where $\mm e\in\mathbb R^{\ndof\times 1}$ is the force direction vector, which is a unit vector in appropriate coordinates.
Without loss of generality, we require $\hat f\in \mathbb R_{>0}$.
We can then follow that
\ea{
\arg\left( \hat{\mm q}\herm_1\,\mm e \hat f \right) &=& \frac\pi 2 \fk \\
\Rightarrow \quad \arg\left( \mm e\tra \hat{\mm q}_1  \right) &=& -\frac\pi 2 \fp \label{eq:localPhaseResonance}
}
\eref{localPhaseResonance} defines the \emph{local phase resonance}:
The displacement response at the drive point needs to lag $90^\circ$ behind the forcing.
More specifically, we have established that this condition must hold with respect to the fundamental harmonic component of response and forcing.
In practice, this can be achieved using an impedance head (which measures both applied force and acceleration at the drive point), and a phase-locked loop.
The control loop is illustrated in \fref{phaseResonance}a.
The phase-locked loop generates a sinusoid and feeds this to the exciter.
The control loop adjusts the oscillation frequency $\Omega$ of this sinusoid until phase resonance is reached.
Of course, if velocity or acceleration are measured instead of the displacement, the target phase must be adjusted from $-90^\circ$ to $0^\circ$ or $+90^\circ$, respectively.
To track the backbone curve, it is common to vary $\hat f$, for instance starting from a certain lower bound and stepwise increase to a certain upper bound \cite{Scheel.2018}.

\subsection{Extraction of modal properties under phase-resonant external forcing}
Once the steady-state response has been recorded for a sequence of points along the backbone curve, the next step is to extract the modal properties.
Per design of the method, the modal frequency is simply the oscillation frequency, which is actually an output of the phase-locked loop.
The Fourier coefficients $\lbrace \hat{\mm q}_0, \hat{\mm q}_1,\ldots \rbrace$ of the periodic modal motion can be obtained via discrete Fourier transform of the recorded sensor data.
The damping ratio can be followed from \eref{PRTone},
\ea{
D = \frac{\real{\hat{\mm q}\herm_1\hat{\mm f}_1}}{2\omega^2 a^2} = \frac{P_1}{\omega^3 a^2}\fk \label{eq:DPRT}
}
where $P_1$ is the period-averaged (active) power supplied by the fundamental harmonic component of the applied excitation force \cite{Scheel.2018}.
In this sense, the damping ratio can be said to follow from the balance between power dissipated within the structure and that supplied by the excitation (to maintain periodic oscillations).
Taking the real part in \eref{DPRT} is consistent with the definition of the active power.
If the phase resonance condition in \eref{PRTtwo} is exactly met, the argument is purely real, so that taking the real part makes no difference.
In experimental practice, the condition is only satisfied with finite precision, so that taking the real part is important to ensure that $D$ is real.
\\
It is crucial to note that the evaluation of $D$ in \eref{DPRT} still requires the value of $a$, the amplitude associated with the mass-normalized mode.
According to \erefs{massnorm} and \erefo{modeShape}, we obtain $a$ from the modal mass
\ea{
a &=& \sqrt{\hat{\mm q}\herm_1~\mm M~\hat{\mm q}_1}\fk \label{eq:aPRTone}\\
&\approx&  \|\mm\Phi_{\mathrm{lin}}^+~\hat{\mm q}_1\| \fp \label{eq:aPRT}
}
To estimate the modal mass in a model-free way, it is proposed to use the condition $\mm\Phi\herm_{\mathrm{lin}}\mm M \mm\Phi_{\mathrm{lin}}=\eye$ where the matrix $\mm\Phi_{\mathrm{lin}}$ contains the mass-normalized linear mode shapes as columns and $\eye$ is the identity matrix.
This yields \eref{aPRT}, where $\square^+$ denotes the Moore-Penrose pseudo-inverse \cite{Scheel.2018}.
The mass-normalized linear mode shapes are determined using conventional Linear Experimental Modal Analysis at low vibration levels.
In the case of external forcing via attached exciters, it is common to use low-level (random) broadband excitation; impact hammer modal testing is an alternative.
$a$ is not only required to determine $D$, but it is also needed to determine the mass-normalized mode shape $\vv$ from the measured (and post-processed) $\hat{\mm q}_1$ via \eref{modeShape}.
\fig[t]{phaseResonance}{Phase-resonant testing using feedback control: (a) excitation by an applied external force, (b) base excitation}{0.9}
\\
The procedure of Nonlinear Experimental Modal Analysis under external forcing by a concentrated phase-resonant load (as in the case of shaker-stinger excitation) can be summarized as:
\begin{enumerate}
  \item Do Linear Experimental Modal Analysis to determine $\mm\Phi_{\mathrm{lin}}$.
  \item Track a sequence of points along the phase-resonant backbone curve using condition \erefo{localPhaseResonance} to obtain amplitude-dependent $\omega$, $\hat{\mm f}_1$ and $\lbrace\hat{\mm q}_0,\hat{\mm q}_1,\ldots\rbrace$.
  \item Use the results of steps 1 and 2 to determine $a$ via \eref{aPRT}, amplitude-dependent $D$ and $\vv$ via \eref{DPRT} and \eref{modeShape}, repsectively.
\end{enumerate}
Steps 1 and 2 do not have to be carried out in that order.
In experimental practice, a finite number, $\nsens$, of response coordinates is measured.
Consequently, the dimension of the vectors $\hat{\mm f}_1$, $\mm e$, $\hat{\mm q}_1$, $\vv$ is $\nsens$; the dimension of the matrix $\mm\Phi_{\mathrm{lin}}$ is $\nsens$ by $\nmod$ where $\nmod$ is the number of retained modes.
The retained modes should span the relevant response frequency band, and one should always use at least as many sensors as modes, $\nsens\geq\nmod$.
%

\section{Theoretical extension of Nonlinear Experimental Modal Analysis to base excitation\label{sec:extension}}
In this section, we closely follow the derivation in \sref{shakerstinger} but consider excitation by a moving base/support instead of the excitation by an applied force.
The base is assumed to have an imposed (one-directional) translational displacement $q_{\mathrm b}(t)$ with known explicit time dependence.
It is useful to consider the coordinates $\mm q$ relative to the base.
The absolute displacement is $\mm q + \mm b q_{\mathrm b}(t)$, where $\mm b\in\mathbb R^{\ndof\times 1}$ is a constant vector.
In appropriate coordinates, $\mm b$ is Boolean with entry one if the corresponding coordinate is aligned with the base motion and zero if it is orthogonal.
If, for instance, all coordinates within $\qq$ are displacements in the same direction as $q_{\mathrm b}$, then all elements of $\mm b$ are $1$.
Consider a mechanical system under base excitation described by the second-order differential equations of motion,
\ea{
\mm M \ddot{\mm q} + \mm g\left(\mm q,\dot{\mm q}\right) = - \mm M \mm b \ddot q_{\mathrm b} \fp \label{eq:EoMbase}
}
Herein, $\mm g$ is the vector of generalized stiffness and damping forces (containing both linear and nonlinear terms in general), and overdot denotes derivative with respect to time $t$.
The generalized mass matrix $\MM$ and generalized force vector $\mm g$ pertain to the situation of constrained base ($q_{\mathrm b}=0$).
The term on the right-hand side can be interpreted as a distributed inertia loading.

\subsection{Phase-resonant base excitation}
Analogous to \sref{shakerstinger}, we require steady-state and resonant conditions.
Hence, \eref{PRTone} and \eref{PRTtwo} still hold.
The fundamental harmonic, $\hat{\mm f}_1$, of the excitation force in these equations can be followed from \eref{EoMbase} as $\hat{\mm f}_1 = \Omega^2\mm M \mm b \hat q_{\mathrm b}$.
Without loss of generality, we require that $\hat q_{\mathrm b}\in\mathbb R_{>0}$.
The phase condition can thus be specified as
\ea{
\arg\left( \hat{\mm q}\herm_1 \mm M \mm b \right) = \frac\pi 2\fp \label{eq:PhaseConditionBase}
}
In the absence of internal resonances and under light damping, the fundamental harmonic of the modal deflection is in good approximation synchronous.
Then, any coordinate aligned with the base motion can be used to check the phase condition between $\mm M \mm b$ and $\hat{\mm q}_1$ locally.
Let's define $\mm e\tra \mm q$ as this considered coordinate where $\mm e\in\mathbb R^{\ndof\times 1}$ is now the selection vector.
If the coordinate is aligned with the base motion in the sense that $\mm e\tra\mm M\mm b>0$, then the local phase resonance condition can be expressed as
\ea{
\arg\left( \mm e\tra \hat{\mm q}_1 \right) = -\frac\pi 2\fp \label{eq:localPhaseResonanceBase}
}
This is identical to \eref{localPhaseResonance}.
The main difference is that we need to use the base motion $q_{\mathrm b}$ instead of the applied forcing $\mm e\tra \mm f$ as reference for the phase condition (\fref{phaseResonance}b).
While the drive point displacement is a natural choice for the formulation of the phase condition in the case of excitation by a concentrated force (local phase resonance), one has to choose an appropriate coordinate in the case of base excitation.
Thanks to the diagonal dominance and positive definiteness of $\mm M$, it should be valid to use any coordinate in the same direction as $q_{\mathrm b}$ (or at least enclosing a positive angle) to satisfy the criterion $\mm e\tra\mm M\mm b>0$.
Throughout the virtual and physical benchmarks analyzed in this work, we did not encounter any cases of highly non-synchronous modal deflection.
However, if the modal deflection should be highly non-synchronous, one could measure the response at multiple locations and approximate the argument in \eref{PhaseConditionBase} using a model-based or model-free estimate of the mass matrix $\mm M$.
We discuss such approximations of the mass matrix in the next subsection, as they are relevant for estimating the modal damping ratio.

\subsection{Identification of modal properties under base excitation}
The modal frequency $\omega$ and the Fourier coefficients, $\lbrace\hat{\mm q}_0,\hat{\mm q}_1,\ldots\rbrace$, of the modal oscillation are determined as in the case of excitation by external forcing (output of phase-locked loop and discrete Fourier analysis, respectively).
To determine the modal damping ratio, we take the magnitude on both sides of \eref{PRTone} and substitute $\hat{\mm f}_1 = \Omega^2\mm M \mm b \hat q_{\mathrm b}$,
\ea{
2D\omega^2 a &=& \omega^2\left|\vv\herm\,\mm M \mm b \hat q_{\mathrm b}\right|\fk \\
2D a &=& \left|\vv\herm\,\mm M \mm b \hat q_{\mathrm b}\right|\fk \label{eq:DbaseOne} \\
D &=& \frac12 \frac{\left|\hat{\mm q}\herm_1\,\mm M \mm b \hat q_{\mathrm b}\right|}{\hat{\mm q}\herm_1\,\mm M\hat{\mm q}_1}\fk \label{eq:DbaseTwo}
}
where we have used that $\Omega=\omega$ along the backbone curve, and we also used the relations $\left|a\vv\right| = \left|\hat{\mm q}\right|$ and $a^2 = \hat{\mm q}\herm_1\,\mm M\hat{\mm q}_1$.
\\
It is useful to remark that both \eref{DPRT} and \eref{DbaseTwo} can be interpreted as balance between power supplied by the excitation force and the dissipation, in average per vibration period.
In the case of base excitation, the excitation force is a distributed inertia loading.
The simple idea of power balance can be found in many studies, and is actually used in \cite{Colin.2020} to estimate damping of a structure under base excitation.
The analysis in \cite{Colin.2020} is strictly limited to a pseudo-quadratic damping term and assumes that the mode shape is amplitude-constant.
In contrast, the theory in the present work makes no assumption on the location/distribution and mathematical form of linear/nonlinear damping, and the mode shape is allowed to change with amplitude.
\\
One can also recognize a similarity to the common definition of equivalent damping, $D = \Delta W_{\mathrm{diss}} / (4\pi E_{\mathrm{pot}}^{\max})$, where $\Delta W_{\mathrm{diss}}$ is the dissipated work per vibration period and $E_{\mathrm{pot}}^{\max}$ is the maximum potential energy reached during a vibration period \cite{Ungar1962}.
It is easy to show that \eref{DPRT} and \eref{DbaseTwo} are consistent with this definition under the additional restriction of synchronous motion and amplitude-invariant mode shape.
The expressions in \eref{DPRT} and \eref{DbaseTwo} are also valid for non-synchronous motion (non-trivial phase lags among the coordinates) and amplitude-dependent mode shape, in full accordance with Single-Nonlinear-Mode Theory.
\\
\eref{DbaseTwo} is not suited for immediate experimental implementation:
First, it involves the mass matrix, which requires an appropriate model (which may not be available).
Second, even if an accurate mass matrix is available, the response would have to be measured at all associated coordinates (which is impractical or even impossible).
In the following, we propose two viable experimental implementations, a model-based variant using \eref{DbaseTwo} as point of departure, and a model-free variant departing from \eref{DbaseOne}.

\subsubsection{Model-based identification of $D$}
The model-based variant uses the linear modes of vibration to evaluate the inner products with the mass in numerator and denominator on the right-hand side of \eref{DbaseTwo}.
We use the approximation,
\ea{
\hat{\mm q}_1 \approx \mm\Phi_{\mathrm{lin}} \hat{\mm \eta}_1\fp
}
Recall that the matrix $\mm\Phi_{\mathrm{lin}}$ contains a truncated set of mass-normalized linear mode shapes as columns.
$\hat{\mm\eta}_1\in\mathbb C^{\ndof\times 1}$ is the vector of fundamental Fourier coefficients of the corresponding modal coordinates.
With this approximation, we obtain from \eref{DbaseTwo}
\ea{
D = \frac12 \frac{\left|\hat{\mm\eta}\herm_1 \, \mm{\Phi}\herm_{\mathrm{lin}} \, \mm M \mm b \hat{q}_{\mathrm b}\right|}{\hat{\mm\eta}\herm_1 \, \hat{\mm\eta}_1} \fk \label{eq:DbaseModelBased}
}
where we have again used that $\mm{\Phi}\herm_{\mathrm{lin}} \mm M \mm{\Phi}_{\mathrm{lin}} = \eye$.
The quantities $\mm M\mm b$, and $\mm{\Phi}_{\mathrm{lin}}$ are to be determined by an appropriate mathematical (\eg finite element) model.
The vector $\hat{\mm \eta}_1$ is to be estimated based on response measurements.
Therefore, it now becomes essential to distinguish between measured and modeled coordinates.
Suppose that we measure a set of $\nsens$ response coordinates and have a mathematical model with $\ndof\geq\nsens$ degrees of freedom.
Then we can relate measured coordinates, modeled coordinates and linear modal coordinates by
\ea{
\hat{\mm q}_{\mathrm{sens},1} = \mm B_{\mathrm{sens}}\hat{\mm q}_1 \approx \mm B_{\mathrm{sens}}\mm{\Phi}_{\mathrm{lin}} \hat{\mm \eta}_1 \fp \label{eq:qsens}
}
Herein, the vector dimensions are $\hat{\mm q}_{\mathrm{sens},1}\in\mathbb C^{\nsens\times 1}$, $\hat{\mm q}_1\in\mathbb C^{\ndof\times 1}$, and $\hat{\mm \eta}_1\in\mathbb C^{\nmod\times 1}$, and the matrix dimensions are $\mm B_{\mathrm{sens}} \in \mathbb R^{\nsens\times\ndof}$, and $\mm{\Phi}_{\mathrm{lin}}\in\mathbb R^{\ndof\times\nmod}$.
From \eref{qsens}, we can derive an estimate for $\hat{\mm\eta}_1$:
\ea{
\hat{\mm\eta}_1 &=& \left(\mm B_{\mathrm{sens}}\mm{\Phi}_{\mathrm{lin}}\right)^+ \hat{\mm q}_{\mathrm{sens},1} \fk \label{eq:etaEstOne}\\
&=& \mm{\Phi}^+_{\mathrm{lin}} \mm B^+_{\mathrm{sens}} \hat{\mm q}_{\mathrm{sens},1}\fk \\
&=& \mm{\Phi}\herm_{\mathrm{lin}} \mm M \mm B^+_{\mathrm{sens}} \hat{\mm q}_{\mathrm{sens},1} \fp \label{eq:etaEst}
}
%
Without loss of generality, we can use a coordinate system in the model so that $\mm B_{\mathrm{sens}}$ contains a unique set of columns of the identity matrix (real-valued).
For $\ndof\geq\nsens$ we can then simplify $\mm B^+_{\mathrm{sens}}$ to $\mm B^+_{\mathrm{sens}} = \mm B\tra_{\mathrm{sens}}\left(\mm B_{\mathrm{sens}} \mm B\tra_{\mathrm{sens}}\right)\inv = \mm B\tra_{\mathrm{sens}}$.
This, along with the explicit expression of the pseudo-inverse of $\mm{\Phi}_{\mathrm{lin}}$ makes the evaluation of \eref{etaEst} simpler than \eref{etaEstOne} for large $\nsens$ (\eg if a finely resolved grid of virtual sensors is measured using a scanning laser-Doppler vibrometer or Digital Image Correlation).
As in the case of external forcing, we should have $\nsens\geq\nmod$ and the retained modes should span the relevant response frequency band.
\\
It should be emphasized that the only ingredients of the mathematical model are the mass distribution (described by $\mm M$) and a set of mass-normalized mode shapes ($\mm{\Phi}_{\mathrm{lin}}$).
In particular, no a priori knowledge on any nonlinearities is needed.
These are rather weak requirements which should be easy to achieve in practice.

\subsubsection{Model-free identification of $D$}
The model-free variant avoids the estimation of linear modal coordinates ($\hat{\mm \eta}_1$) and the mass matrix altogether.
This is achieved by exploiting the relation
\ea{
\vv\herm\mm M\vv &=& 1 \fk \\
\vv\herm\mm M\hat{\mm q}_1 &=& a\ee^{\ii\theta} \fk \\
\vv\herm\mm M &\approx& a\ee^{\ii\theta} \hat{\mm q}^+_1 \fk \\
\vv\herm\mm M &\approx& a\ee^{\ii\theta} \frac{\hat{\mm q}\herm_1}{\hat{\mm q}\herm_1 \hat{\mm q}_1}\fp \label{eq:phiMinv}
}
The explicit expression of the pseudo-inverse in the last step holds for any vector $\hat{\mm q}_1 \neq \mm 0$.
Substituting \eref{phiMinv} into \eref{DbaseOne} yields
\ea{
D \approx \frac12 \frac{\left| \hat{\mm q}\herm_1 \mm b \hat q_{\mathrm b} \right|}{ \hat{\mm q}\herm_1 \hat{\mm q}_1} \fp \label{eq:DbaseModelFree}
}
%
To improve the approximation, in particular for small $\nsens$, it is useful to consider the spatially continuous representation of the inner products in the numerator and the denominator on the right-hand side of \eref{DbaseModelFree},
\ea{
D \approx \frac12 \frac{\left|
\int\limits_{\mathcal B} \vec{\hat q}_1^* \cdot \vec b \,\dd V \, \hat q_{\mathrm b}
\right|}{
\int\limits_{\mathcal B} \vec{\hat q}_1^* \cdot \vec{\hat q}_1\, \dd V
} \fk \label{eq:DbaseModelFreeContinuous}
}
where $\square^*$ denotes complex-conjugate and the vector fields $\vec{\hat q}_1$ and $\vec b$, defined continuous over the whole body $\mathcal B$, are used instead of the spatially discrete counterparts $\hat{\mm q}_1$ and $\mm b$, respectively.
Accordingly, the inner products of the column vectors in \eref{DbaseModelFree} are replaced by the inner products of the vector fields in \eref{DbaseModelFreeContinuous}.
To implement \eref{DbaseModelFreeContinuous} using a discrete set of coordinates, a quadrature rule is needed.
Suppose that we have a one-dimensional continuum (\eg a beam) and place the sensors equidistantly.
Using \eref{DbaseModelFree} directly corresponds to applying the rectangular rule to the integrals in \eref{DbaseModelFreeContinuous}.
Better convergence can be achieved with a more sophisticated quadrature rule.
For different quadrature rules, convergence with the number of sensors is analyzed in \sref{conv}.

\subsubsection{Remarks on the mass-normalization of the modal deflection shape}
It should be noted that neither the mass-normalized modal deflection shape, $\vv$, nor the corresponding modal amplitude, $a$, are needed to determine $D$ in the case of base excitation, as opposed to the case of excitation by external forcing.
To set up the nonlinear modal oscillator model in \eref{NMROM}, one generally needs the mass-normalized modal deflection shape $\vv$.
In the special case where the response to base excitation is requested, one can use again the relation $\vv\herm \mm M/(a\ee^{\ii\theta}) = \hat{\mm q}\pinv_1$ to eliminate $a$ and $\vv$ within the base excitation term.
Then, the quantity $\hat{\mm q}\pinv_1\mm b$ appears as a factor on the right hand side, which one may use as amplitude-like parameter (actually its reciprocal is a measure for the vibration level).
In general, the mass normalization can be done, again, model-based or model-free.
The model-free variant can be carried out analogous to the case of excitation by external forcing using \eref{aPRT}.
To this end, one has to determine $\mm\Phi_{\mathrm{lin}}$ experimentally.
In contrast to the case of shaker-stinger excitation, the applied force cannot be measured directly, and hence the common approach to do shaker testing and carry out the mass normalization via the drive-point frequency-response function is infeasible.
An alternative is to use impact hammer modal testing.
If the model-based variant is pursued to identify $D$ it seems natural and mathematically consistent to carry out the mass normalization based on the model, too.

\subsubsection{Summary of procedures for Nonlinear Experimental Modal Analysis under base excitation}
The procedure of the model-based variant can be summarized as:
\begin{enumerate}
  \item Setup a linear (\eg FE) model and do a linear modal analysis to determine $\mm{\Phi}\herm_{\mathrm{lin}} \, \mm M \mm b$ and $\mm B_{\mathrm{sens}}\mm{\Phi}_{\mathrm{lin}}$.
  \item Track a sequence of points along the phase-resonant backbone curve using condition \erefo{localPhaseResonanceBase} to obtain amplitude-dependent $\omega$, $\hat q_{\mathrm b}$ and $\lbrace \hat{\mm q}_{\mathrm{sens},0},\hat{\mm q}_{\mathrm{sens},1},\ldots\rbrace$.
  \item Use the results of steps 1 and 2 to estimate $\hat{\mm \eta}_1$ via \eref{etaEstOne}, determine $D$ via \eref{DbaseModelBased}. \emph{Optional:} Determine $a = \sqrt{\hat{\mm \eta}\herm_1\hat{\mm\eta}}$ and $\vv$ via \eref{modeShape}.
\end{enumerate}
Steps 1 and 2 do not have to be carried out in the given order.
The sub-step within step 3 is optional in the sense that it is only required if mass-normalized modal deflection shapes are needed.
Of course, model and experimental setup must be consistent and it should be ensured that $\nmod\leq\nsens\ll\ndof$.
\\
The procedure of the model-free variant can be summarized as:
\begin{enumerate}
  \item Track a sequence of points along the phase-resonant backbone curve using condition \erefo{localPhaseResonanceBase} to obtain amplitude-dependent $\omega$, $\hat q_{\mathrm b}$ and $\lbrace \hat{\mm q}_{\mathrm{sens},0},\hat{\mm q}_{\mathrm{sens},1},\ldots\rbrace$.
  \item Use the results of step 1 to determine amplitude-dependent $D$ via \eref{DbaseModelFreeContinuous}.
  \item \emph{Optional:} Do Linear Experimental Modal Analysis to determine $\mm\Phi_{\mathrm{lin}}$ and use the results of step 1 to obtain $a$ via \eref{aPRT} and $\vv$ via \eref{modeShape}.
\end{enumerate}
Step 1 is the same as step 2 in the model-based variant.
Step 3 is optional in the sense that it is only required if mass-normalized modal deflection shapes are needed.
Steps 2 and 3 do not have to be carried out in the given order.

\subsubsection{Validation of model-based and model-free estimation of $D$ for the linear case}\label{sec:conv}
%
\fig[h!]{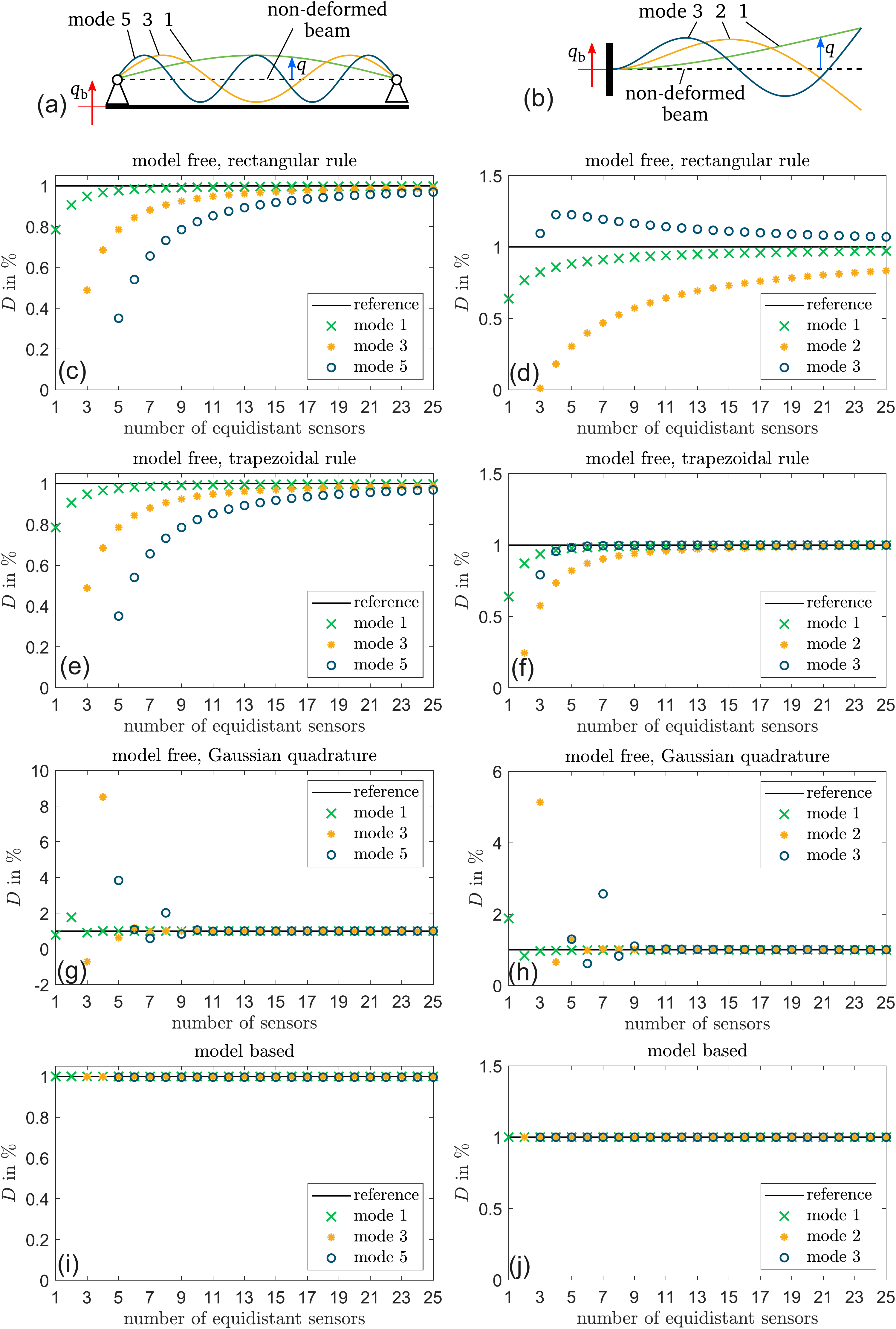}{Analysis of the convergence of the estimated modal damping ratio with the number of sensors: (first row) definition of one-dimensional problem setting and illustration of the modal deflection shapes, (second row) model-free using rectangular rule with equidistant sensors, (third row) model-free using trapezoidal rule with equidistant sensors, (fourth row) model-free using Gaussian quadrature rule, (last row) model-based
}{0.75}
%
We consider two problem settings involving a one-dimensional continuum respectively, a pinned-pinned and a cantilevered Euler-Bernoulli beam (\fref{theory}).
Only the odd modes are considered in the case of the pinned-pinned beam since the even modes are orthogonal to the base excitation.
A constant modal damping ratio of $1\%$ is specified for all modes.
As the problem is linear, the steady-state response to harmonic excitation with the respective natural frequency is simply expressed in closed form in the frequency domain.
The pinned/clamped points have zero deflection ($q=0$) per definition.
The trivial response at these points is used in the identification, but these points are not counted as sensors.
One can see excellent accuracy in the case of the model-based variant.
It merely needs to be ensured that the number of sensors is at least equal to the number of analyzed modes, $\nsens\geq\nmod$ (otherwise the modes cannot be distinguished).
Compared to the model-based variant, the model-free variant requires more sensors.
For equidistant discretization, trapezoidal and rectangular rule differ merely by the weighting of the points on the boundaries.
Thus, there is no difference if all boundaries are pinned or clamped.
However, if one of the structure's boundaries is free, as in the case of the cantilevered beam, the trapezoidal rule converges much more quickly.
In the considered cases, using $3$ to $5$ times as many equidistant sensors as the mode order in the trapezoidal rule provides high accuracy.
High accuracy can also be achieved using Chebyshev-Gauss quadrature.
Compared to trapezoidal quadrature, the errors are larger for very few sensors, and then decrease more rapidly with the number of sensors.
It should be remarked that Gaussian quadrature does not exploit the known trivial response at clamped/pinned boundaries.
Moreover, in experimental practice, it may be difficult to place the sensors at the Gauss points and it may be much easier to implement a more regular grid (\eg using a scanning laser-Doppler vibrometer or Digital Image Correlation).
Therefore, we use the trapezoidal rule with (almost) equidistant sensors in the remainder of this article.
\\
It should be emphasized that many imperfections encountered in experimental practice are deliberately not accounted for here (\eg sensor or process noise, model error, exciter-structure interaction) and the considered problem is linear.
These aspects are analyzed separately via virtual and physical experiments in \srefs{resultsVirtual} and \srefo{resultsPhysical}.

\section{Results of the virtual experiments\label{sec:resultsVirtual}}
In this section, we validate and assess the proposed methods for two virtual experiments.
Subsequently, the results of a physical experiment are presented in \sref{resultsPhysical}.
An important advantage of virtual experiments over physical ones is that a clear reference is available (EPMC).
The first virtual experiment consists of a cantilevered beam with an elastic dry friction (Jenkins) element,
the second one of a clamped-clamped beam subjected to nonlinear bending-stretching coupling.
As can be followed from the theoretical development, a crucial aspect of the methods is the number and placement of the sensors.
This is because the sensors are not only used as reference for the feedback control of the excitation, but they are used to estimate the modal contributions (model-based variant), or the vibration energy and the power provided by the distributed inertia loading (denominator and numerator on right-hand side of \eref{DbaseModelFreeContinuous} in model-free variant).
Three different sets of equidistantly spaced sensors are considered for each virtual experiment (\tref{sensorSets}).
The index of the sensor set indicates the number of sensors; \eg, the sensors set $\mathcal S_5$ contains $5$ sensors.
The sensor numbers are defined in the top-left of \fref{VEfriction} and \fref{VEgeomNL}, respectively.
\tab[h]{l|ccc}{
benchmark & 1 sensor & 2 sensors & all sensors\\
\hline
cantilevered beam & $\mathcal S_1 = \lbrace 6 \rbrace$ & $\mathcal S_2 = \lbrace 3,6 \rbrace$ & $\mathcal S_6 = \lbrace 1,2,3,4,5,6 \rbrace$ \\
clamped-clamped beam & $\mathcal S_1 = \lbrace 3 \rbrace$ & $\mathcal S_2 = \lbrace 2,4 \rbrace$ & $\mathcal S_5 = \lbrace 1,2,3,4,5 \rbrace$
}{Sensors sets used for the virtual and physical experiments; the sensor locations are defined in the top-left of \fref{VEfriction} and \fref{VEgeomNL}}{sensorSets}

\subsection{Virtual experiment with a friction-damped beam\label{sec:resultsVirtualFriction}}
A schematic illustration of the problem setting is given in \fref{VEfriction}a.
Setup and properties are similar to the Rubbing Beam Resonator (RubBeR) developed in \cite{Scheel.2020b}.
The cantilevered beam was modeled according to the Euler-Bernoulli theory and had the following properties: bending stiffness $EI = 12,49~\mathrm{Nm}^2$, mass per unit length $\rho A = 7.047~\mathrm{kg/m}$, length $L = 0.7~\mathrm{m}$.
The continuous model was truncated to the $5$ lowest-frequency bending normal modes (without friction element), and it was ensured that the depicted results do not change if the truncation order is further increased.
Linear modal damping of $1\%$ was added to all retained modes.
The elastic dry friction element was placed at $3/7L$ from the clamping, and its parameters were: friction limit force $\mu N = 1~\mathrm{N}$, stiffness $k_{\mathrm t} = 27.47~EI/L^3$.
\\
The amplitude-dependent modal frequency and damping ratio of the lowest-frequency mode are depicted in \fref{VEfriction}b-d.
As amplitude, the fundamental harmonic magnitude of the tip deflection $\hat w(L)$, normalized by the beam length $L$, is considered.
The nonlinear modal frequency $\omega$ is normalized by the linear one $\omega_0$.
The Extended Periodic Motion Concept serves as reference \cite{Krack.2015a}.
More specifically, the nonlinear modal analysis was carried out by solving $\mm M \ddot{\mm q} - 2D\omega \mm M\dot{\mm q} + \mm g\left(\mm q, \dot{\mm q}\right) = \mm 0$ using (multi-)Harmonic Balance and numerical path continuation. 
To this end, the open source Matlab tool NLvib was used \cite{Krack.2019}.
The harmonic truncation order was set to $H=7$ and it was ensured that the depicted results do not change by a further increase of $H$.
In the virtual experiment, a phase-locked loop was used to ensure phase resonance between the base acceleration, $\ddot q_{\mathrm b}$, and sensor 1.
The exciter was modeled as ideal, \ie, the output of the controller is directly used as base acceleration.
%
\fig[t]{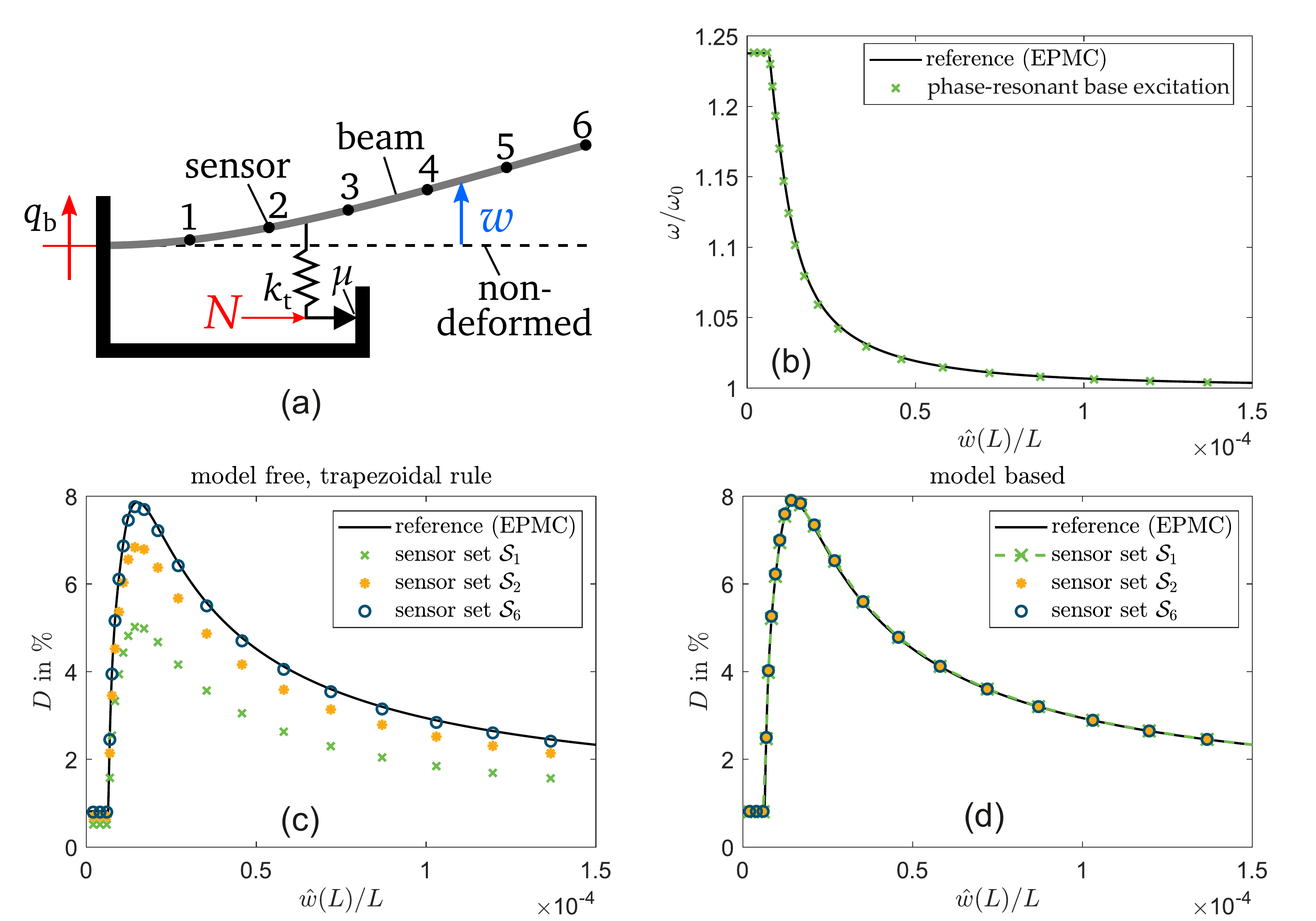}{
Virtual experiment with a friction-damped beam: (a) definition of problem setting, (b) modal frequency, (c)-(d) modal damping ratio estimated by model-free and model-based method, respectively
}{1.0}
\\
The modal frequency shift by more than $20\%$ and the damping increase from about $1\%$ to $8\%$ with a subsequent decrease underline the severe nonlinearity of the friction-damped beam.
The phase-locked loop robustly achieves phase resonance in the depicted amplitude range.
This leads to an excellent agreement of proposed method (phase-resonant base excitation) and reference (EPMC) in terms of amplitude-dependent modal frequency (\fref{VEfriction}b).
The model-based variant of the method achieves excellent agreement with respect to the amplitude-dependent damping already for a single sensor.
It should be remarked that in this case, the fundamental harmonic of the modal deflection shape is dominated by the corresponding linear one (although there is a substantial contribution of higher harmonics).
Compared to the model-based variant, the model-free variant requires more sensors.
For a number of $6$ sensors, the estimated modal damping curve is in excellent agreement with the reference.
In this case, the trapezoidal rule was used for the evaluation of \eref{DbaseModelFreeContinuous}.
The convergence with the number of sensors is consistent with the results for the linear case in \sref{conv}.

\subsection{Virtual experiment with a clamped-clamped beam subjected to nonlinear bending-stretching coupling\label{sec:resultsVirtualNLGeom}}
A thin clamped-clamped beam was considered for the second virtual experiment.
The beam is initially straight (in contrast to that in the physical experiment, which is initially curved) and has no pretension (\fref{VEgeomNL}a).
As longitudinal displacement is prohibited by the clamping, bending deformation causes longitudinal loading, which increases the bending stiffness.
This bending-stretching coupling corresponds to a geometric nonlinearity.
The clamped-clamped beam was modeled using simplified beam theory as \eg in \cite{nayf1979} with the following properties: $EI = 0.245~\mathrm{Nm}^2$, $\rho A = 0.1078~\mathrm{kg/m}$, $L=0.14~\mathrm{m}$.
The continuous model was truncated to the $5$ lowest-frequency normal modes, and it was ensured that the depicted results do not change if the truncation order is further increased.
Linear modal damping of $1\%$ was added to all retained modes.
\\
The amplitude-dependent modal frequency and damping ratio of the lowest-frequency mode are depicted in \fref{VEgeomNL}b-d.
As amplitude, the fundamental harmonic magnitude of the beam's center deflection $\hat w(L/2)$, normalized by the beam thickness $h$, is considered.
Otherwise, the analysis is carried out analogous to the first virtual experiment.
\fig[t]{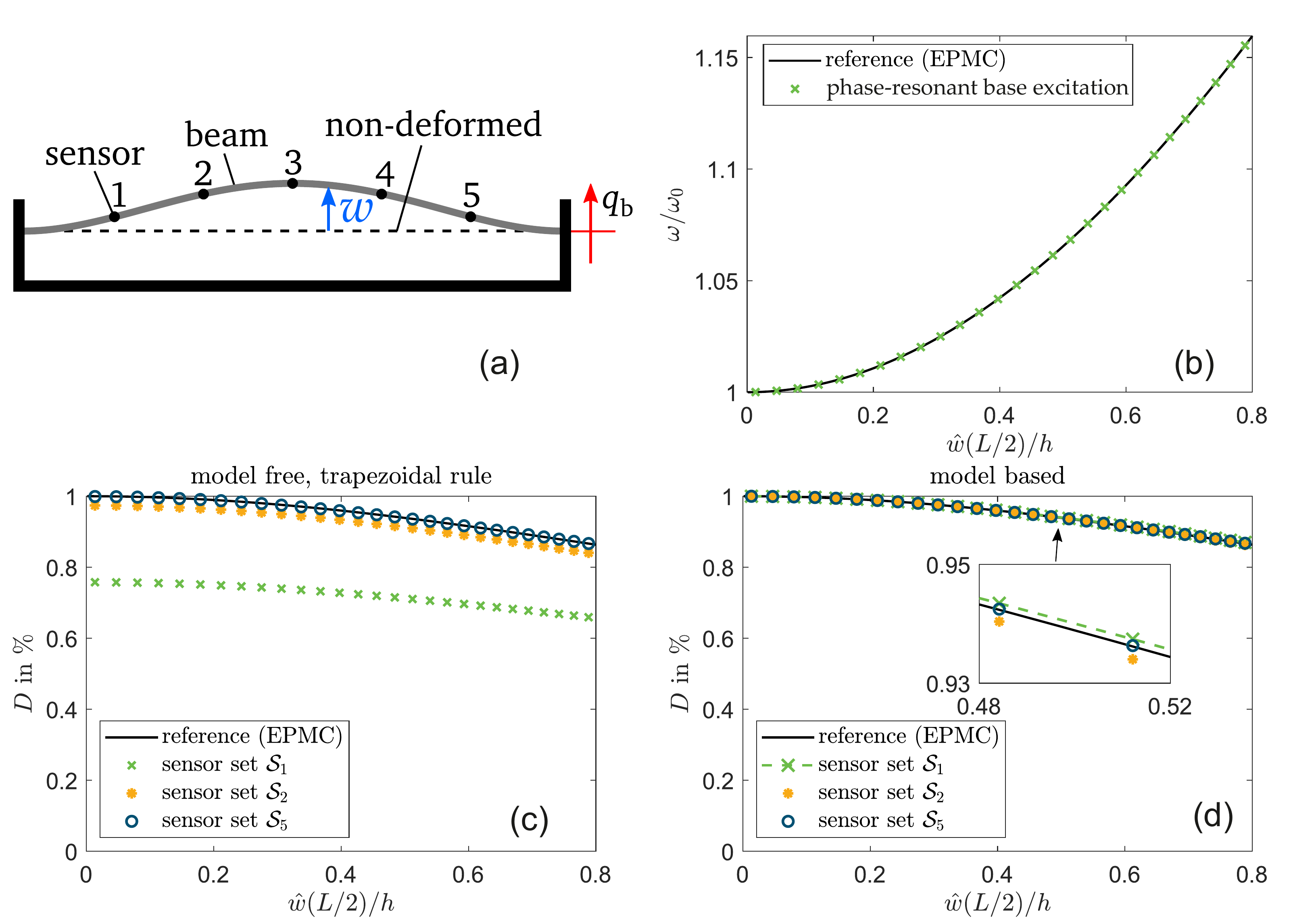}{
Virtual experiment with a clamped-clamped beam subjected to nonlinear bending-stretching coupling: (a) definition of problem setting, (b) modal frequency, (c)-(d) modal damping ratio determined by model-free and model-based method, respectively
}{1.0}
\\
In the considered amplitude range, the geometric hardening nonlinearity leads to a modal frequency shift by more than $15\%$.
The proposed phase-resonance testing achieves excellent agreement with the EPMC reference.
The modal damping stems from the prescribed linear modal damping of $1\%$ for all retained linear normal modes.
The slight amplitude-dependence of the modal damping ratio can be explained by the change of modal frequency.
Again, the model-based damping estimation achieves excellent agreement already for a single sensor.
For this example, the model-free damping estimation yields very good results already for only two sensors.

\section{Results of the physical experiment\label{sec:resultsPhysical}}
As physical experiment, an initially curved beam is considered which is clamped at both ends via bolted joints (\fref{PEproblemsetting}a).
The setup is as in \cite{Abeloos.2022}.
Two physically quite different types of nonlinearity are expected to characterize the dynamics:
The bending-stretching coupling is expected to cause a stiffness nonlinearity, similar to the second virtual experiment analyzed in \sref{resultsVirtualNLGeom}.
Nonlinear damping is expected due to dry frictional interactions within the bolted joints (micro-slip).
The clamped beam is mounted on a frame, which is placed onto a shaker (Brüel \& Kjær vibration exciter Type 4809) via an impedance head.
Although the frame is relatively stiff, compared to the thin beam, it will undergo some elastic deformation, as illustrated in \fref{PEproblemsetting}b.
The motion of the clamping blocks at the sensor positions indicated in \fref{PEproblemsetting}a is considered as base motion.
The base motion is measured using a single-point laser-Doppler vibrometer (SPV2).
The elastic deformation of the beam (relative to the base) is determined using a differential single-point laser-Doppler vibrometer (SPV1, SPV1-REF).
The analog signals of these two vibrometers are fed into a dSPACE MicroLabBox, where the phase-locked loop is implemented.
The controller output is fed to the amplifier (Brüel \& Kjær vibration exciter Type 2718) of the shaker.
The response is recorded at the five points along the beam, simultaneously, using a multi-point vibrometer (MPV).
For the proposed method of modal damping identification, only the response data recorded by the vibrometers is used.
The acceleration and force measured by the impedance head was only used for checking the plausibility of the modal damping ratio, as explained later.
\fig[t]{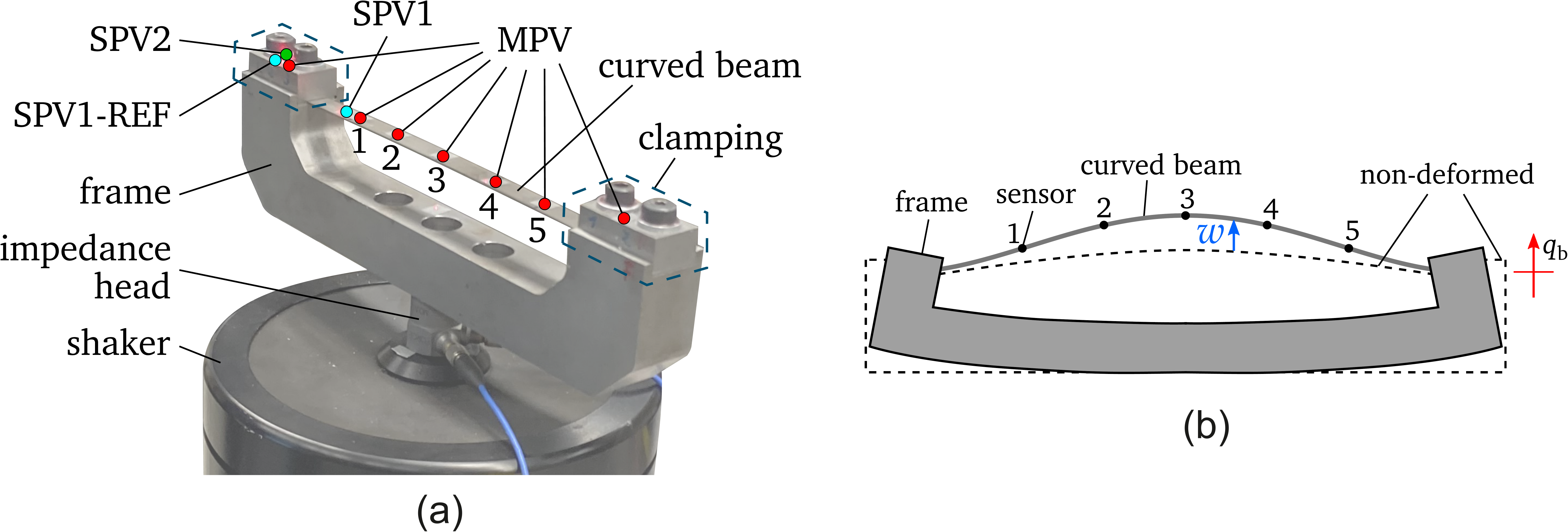}{Physical experiment: (a) illustration of the test rig, (b) schematic illustration of the considered mode shape; SPV1 and SPV1-REF refer to the first single-point laser-Doppler vibrometer, which is used in differential operation. SPV2 refers to the second single-point laser-Doppler vibrometer. MPV refers to multi-point vibrometer.
}{1.0}
\\
The phase-locked loop is implemented using synchronous demodulation, as in \cite{Scheel.2020b}, where the control loops are explained in detail.
The parameters of the phase-locked loop are: initial frequency $\omega_{\mathrm m} = 1,948~\mathrm{rad/s}$, proportional gain $K_{\mathrm p} = 600~1/\mathrm{s}$, integral gain $K_{\mathrm i} = 200~1/\mathrm{s}^2$.
For the frequency response tests (carried out for validation as explained below), a differential gain with $K_{\mathrm d} = 10$ was included in the phase-locked loop.
For these tests, an additional controller was introduced to keep the excitation level constant.
More specifically, a PI controller with proportional gain $K_{\mathrm p}^a = 600$ and integral gain $K_{\mathrm i}^a = 400~1/\mathrm{s}$ was used to control the magnitude of the fundamental harmonic of the base velocity to a set value.
Note that this magnitude is an output of the synchronous demodulation (see \eg \cite{Scheel.2020b} for details).
The output of the PI controller is fed to the gain of the voltage input to the shaker amplifier.
A crucial aspect in shaker-based testing is exciter-structure interaction.
With the described strategy, only the phase (and the magnitude in the frequency response test) of fundamental harmonic of the base motion is controlled.
Throughout the nonlinear backbone and frequency response tests, the distortion factor of the base velocity was greater than 0.95\footnote{Here, the definition of the distortion factor according to \cite{Shmilovitz.2005} is used.}.
This is considered quite good by the authors and therefore a compensation of higher harmonics was deemed unnecessary.
Moreover, based on the system's linear natural frequencies, there is no reason to expect a high structural sensitivity to the small higher harmonics present in the excitation signal.
One reason for the low contribution of higher harmonics is probably that the vibrating mass is low with respect to the total moving mass:
The ratio between the mass of the beam (8.6 gram) and the total moving mass (moving parts of shaker, impedance head, frame, beam, clamping blocks and screws; 1.1kg in total) is less than $1\%$.
As the waveform of the excitation is in very good accordance with the desired purely sinusoidal form of the desired magnitude and phase, we regard the effect of the exciter-structure interaction on the measurement results as negligible.
The excitation level was stepwise increased and then decreased in the backbone test.
The wait time was $5,900$ and $3,100$ excitation periods in the backbone and frequency-response test, respectively, and the hold time was $300$ excitation periods per step.
It is assumed that the transients decay sufficiently during the wait time.
The hold time corresponds to the time span used for the further signal processing in terms of the model-based and model-free method variants.
%
\fig[t]{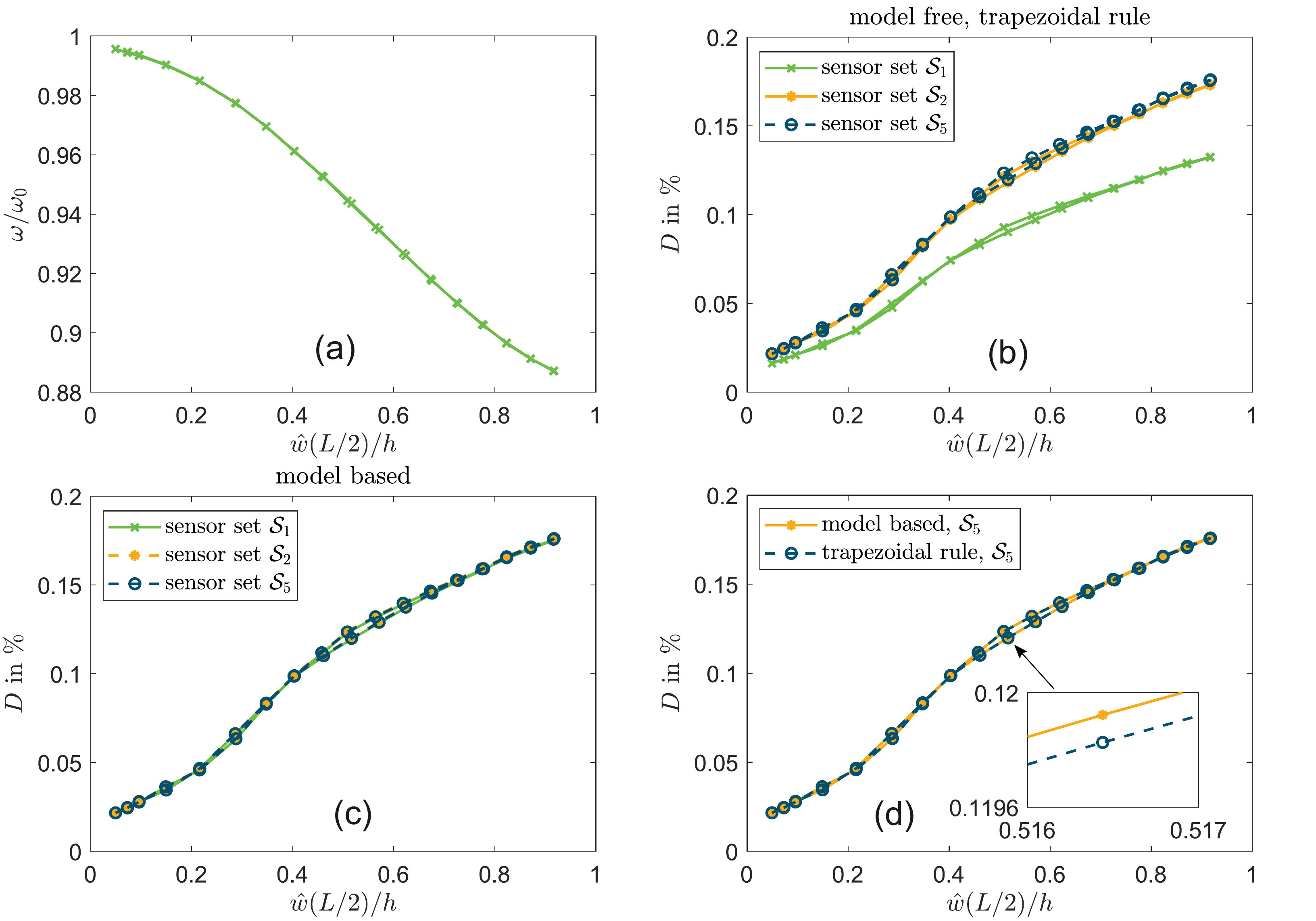}{Results of the physical experiment: (a) modal frequency, (b)-(c) modal damping ratio determined by model-free and model-based method, respectively, (d) direct comparison
}{1.0}
\\
The amplitude-dependent modal frequency and damping ratio of the lowest-frequency mode are depicted in \fref{PEomD}.
As amplitude, the fundamental harmonic magnitude of the beam's center deflection $\hat w(L/2)$, normalized by the beam thickness $h$, is considered.
Due to the initial curvature, the bending-stretching coupling leads to a softening rather than a hardening effect in the considered amplitude range.
Hardening is expected for even higher amplitudes.
The modal frequency shift of about $12\%$ and the damping increase from about $0.02\%$ to about $0.18\%$ correspond to a substantial nonlinearity.
The results of forward and backward stepping along the backbone indicate very good repeatability.
These properties make the test rig well-suited for the assessment of the proposed method.
Taking a closer look at the damping-amplitude curves, a slight difference between forward and backward stepping can be recognized.
Potential physical reasons for this are thermal strains and the non-uniqueness of the tangential preload of the contact interfaces within the bolted connections.
Model-based and model-free method variants perform similarly as in the corresponding virtual experiment (\sref{resultsVirtualNLGeom}).
In particular, both variants agree perfectly for a sufficiently large number of sensors (5 in this case).
The definition of the sensor sets is as in \tref{sensorSets} (last row).
Very robust results can already be achieved with the model-free variant using only 2 sensors, while 1 sensor is sufficient already for the model-based variant.
\fig[t]{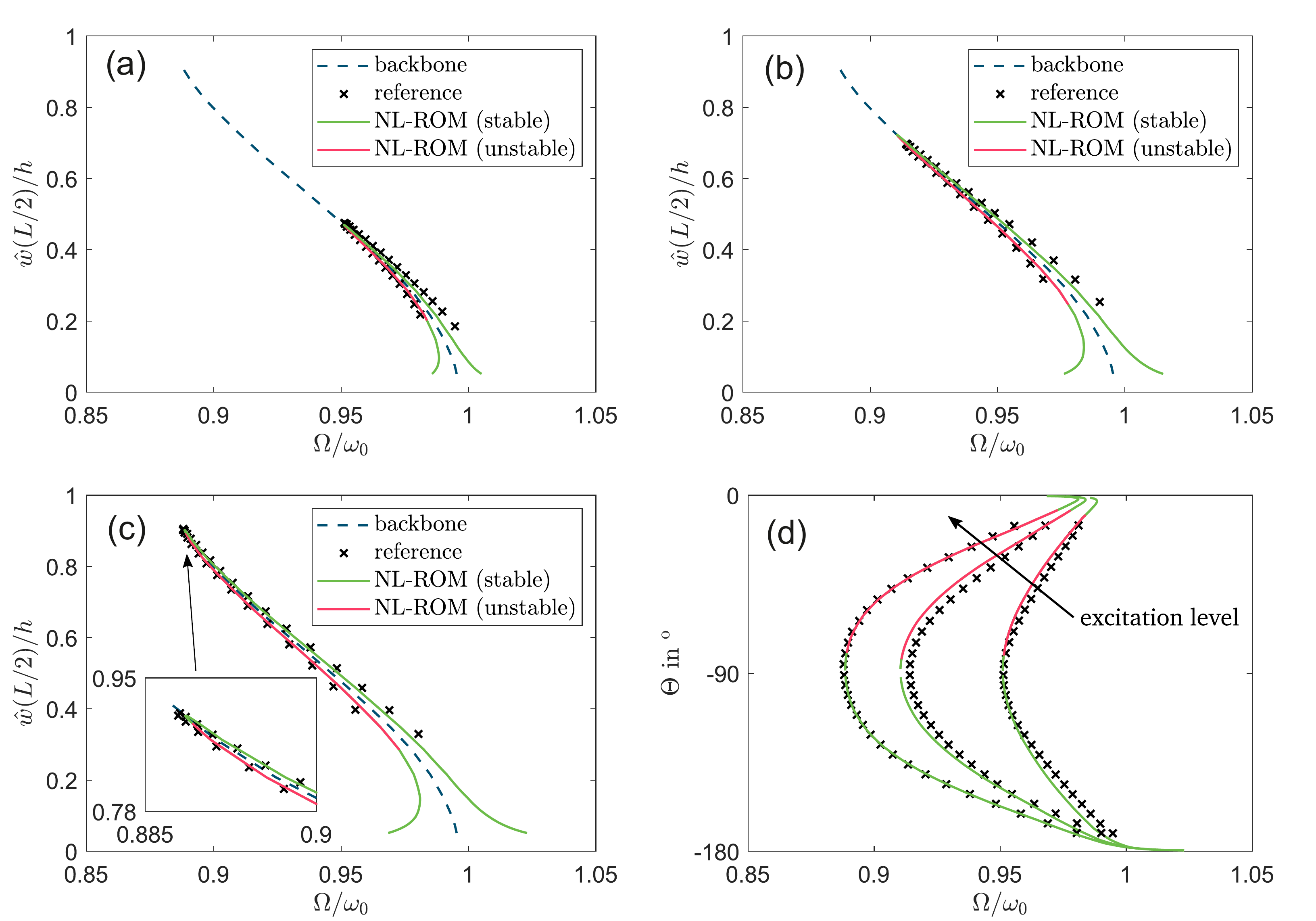}{Results of the physical experiment: (a)-(c) amplitude-frequency curve and backbone curve for three different excitation levels, (d) phase-frequency curves. The definition of the sensor sets is as in \tref{sensorSets}. NL-ROM stands for nonlinear-mode-based predictions.}{1.0}
\\
It should be emphasized that the consistency of both method variants does not imply that the results are valid.
For an appropriate validation, we use Single-Nonlinear-Mode Theory to make predictions of the frequency response using the identified amplitude-dependent modal properties, and we compared against direct measurements.
The results are presented in \fref{PEfreqResp}.
The direct measurements (reference) were obtained, again, using the phase-locked loop (but varying the target phase) and an additional controller to ensure constant base velocity level throughout the frequency response curve.
The predictions based on Single-Nonlinear-Mode Theory (NL-ROM) are in very good agreement with the reference, and the measured backbone indeed connects the maxima of the frequency response curves.
The asymptotic stability of the nonlinear-mode-based results was determined by analyzing the eigenvalues of the Jacobian evaluated at the fixed points of \eref{NMROM} (after splitting into real and imaginary part).
From the high accuracy of the nonlinear-mode-based predictions, we conclude the high accuracy of the amplitude-dependent modal properties (including the damping ratio).
\\
We would like to remark that we first analyzed the test rig in the externally-forced configuration.
Here, the structure under test consists not only of the beam but also the frame.
Moreover, phase resonance needs to be ensured between response and forcing at the drive point (impedance head).
Remarkably, the resulting modal damping ratio was about twice as large for the highest amplitudes considered in this work.
The higher damping ratio can be explained by the dissipation in the (stiff but not rigid) frame.

\section{Conclusions\label{sec:conclusions}}
We developed a method that is, for the first time, able to determine amplitude-dependent modal properties of nonlinear structures under base excitation.
As in the case of shaker-stinger excitation, the method relies on tracking the phase-resonant backbone curve.
Here the only difference is that the base motion rather than the applied force serves as reference for the definition of the resonant phase lag.
A challenge lies in the damping quantification, as the applied force generally cannot be measured under base excitation.
We developed a model-free and a model-based variant which both rely on pure response measurement and thus do not resort to force measurement.
The virtual and physical experiments demonstrate high robustness, high accuracy and reasonably quick convergence with the number of sensors.
Even for the model-free variant, we do not think that the slightly larger number of required sensors will be an important practical limitation.
The model-based variant provides useful results already for a single sensor, and the model requirements are relatively weak, since only the mass distribution and the linear mode shapes are needed.
\\
The authors are convinced that the proposed method will be useful for structures that cannot properly be tested using shaker-stinger excitation but rather base excitation with a large shaker.
In the future, it would be interesting to assess the opportunities and limitations of the method for more complicated geometries and higher-order modes.
It could also be interesting to compare the method against frequency-response-based modal parameter identification \eg based on response-controlled stepped sine testing \cite{Karaagacl.2021}.

\section*{Acknowledgement}
The authors are very grateful to Polytec GmbH, in particular Patric Gehring and Dennis Berft, for providing equipment and support during the experimental studies.
This work was partly funded by the Deutsche Forschungsgemeinschaft (DFG, German Research Foundation) [Project 402813361], and by MTU Aero Engines AG, Germany.
The authors are grateful to MTU for giving permission to publish this work.

\end{document}